\documentstyle[amssymb,aps,twocolumn,epsfig]{revtex}

\begin{document}

\title{Two-proton events in the $^{17}$F$(p,2p)^{16}$O reaction}

\author{
L.\ V.\ Grigorenko$^{1,2,3}$,
R.\ C.\ Johnson$^{1}$,
I.\ J.\ Thompson$^{1}$, and
M.\ V.\ Zhukov$^{4}$.
}

\address{
$^{1}$Department of Physics, University of Surrey, Guildford GU2 7XH, U.K.\\
$^{2}$Russian Research Center ``The Kurchatov Institute'', 123182 Moscow, Russia\\
$^{3}$Gesellschaft f\"{u}r Schwerionenforschung mbH, Planckstr.\ 1,
D-64291 Darmstadt, Germany\\
$^{4}$Department of Physics, Chalmers University of Technology\\
and G\"{o}teborg University, S-41296 G\"{o}teborg, Sweden \\
}

\maketitle

\abstract{In a recent experimental study  \cite{gom01} of the reaction
$^{17}$F$(p,2p)^{16}$O two-proton events were measured from excitations near a
$1^-$, $E^*=6.15$ MeV state in $^{18}$Ne. We calculate by
means of $R$-matrix theory the resonant two-proton production cross
section and branching ratios. We conclude that it is unlikely that
two-proton production via population of the $1^-$ state is sufficient to
explain the observed two-proton events. Alternative sources of such events are
discussed.}

\vspace{0.5mm} \hrulefill \mbox{\quad} \vspace{3mm}

%==============================================================================

{\em Introduction.}
In recent paper \cite{gom01},  states in $^{18}$Ne were populated
in the $^{17}$F+$p$ reaction.
High quality results for the one-proton excitation
function are obtained in the energy range 0.4--2.45 MeV,
and the question arises whether exotic two-proton emission processes
can be seen in such an experiment.
Because there are no intermediate states in $^{17}$F
available for sequential decay, the region of the $^{18}$Ne spectrum below 6.5 MeV
might be envisaged as a good place to study simultaneous two proton decay.
This needs a theory of two proton decays.

We therefore review the theoretical foundations of the two
mechanisms (as mentioned in the introduction of paper \cite{gom01}):
(i) diproton mechanism \cite{gol60} and (ii) democratic decay \cite{boc92}.
In his often-referenced paper \cite{gol60}, Goldansky pointed that a curious
quantum mechanical effect is possible in some proton-rich
nuclei: what he calls ``true two-proton emission'' takes place if there are
{\em no two-body decay channels at all} due to energy conditions in the
subsystems. The fingerprint of this effect should be ``the energy correlation
between the protons during the two-proton decay, which leads to their energies
being almost equal''. Goldansky also noticed that the estimated penetrability for
two $s$-wave protons is very close to the penetrability for the `diproton'
(charge 2 `particle' with zero energy of internal motion). We remark that this
similarity of penetrabilities has led to the widespread idea that Goldansky expected
two-proton decay to produce two protons with almost coincident {\em velocities},
whereas he predicted merely coincident {\em energies}.

A slightly different view on the phenomenon is provided by the concept of
democratic decay \cite{boc84,dan87,boc92}:
it was shown experimentally for decay of the $^6$Be g.s.\
that the energy distribution between the protons is broad (`democratic')
and no `diproton'-type correlation was observed.
It was suggested that the reason for the broad distributions is the absence
of narrow states in all subsystems in this decay. This is less stringent
condition than the condition for `true two-proton decay' in the paper of Goldansky,
namely that the
``positive binding energy of the first proton must be larger than the
half width of the emission of the second one''.
However, the qualitative prediction of Goldansky that protons should
evenly share the energy was confirmed in these studies.
It was shown in \cite{dan87} that the energy distributions in
democratic decays can be described by the expression
\begin{equation}
dN/d\epsilon \propto \sqrt{\epsilon(E_{2p}-\epsilon)} \; |A|^2
\label{dem}
\end{equation}
where $\epsilon$ is relative energy of the protons,
$E_{2p}$ is the energy of the resonance relative to the $2p$ threshold,
and amplitude $A$ depends only weakly on $\epsilon$.

%===============================================================================
\begin{figure}[tb]
\centerline{
        \psfig{figure=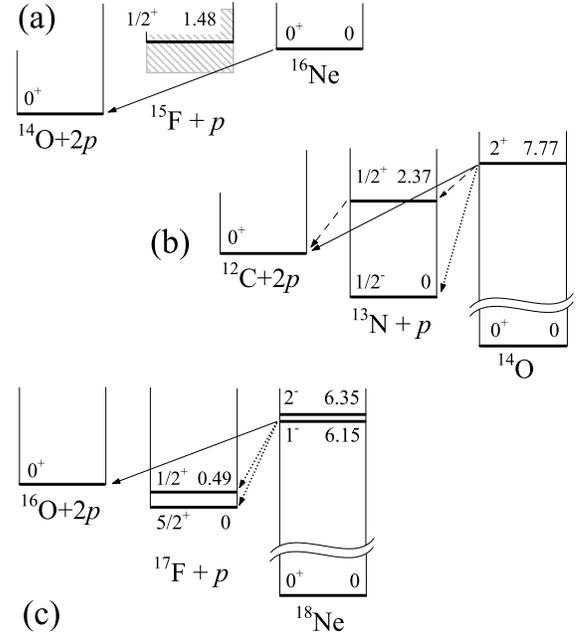,width=0.41\textwidth}
        }
\smallskip

\caption{Schemes of different two-proton decays illustrating (a) true two-proton
decay, (b) sequential decay, and (c) simultaneous emission of two protons, but with
two-body $^{17}$F + $p$ channel dominating.}

\label{sch}

\end{figure}
%===============================================================================

More detailed insight into the different decay modes can be found in the paper
\cite{gri01a}, where the applicability of simple models of two-proton decay is
carefully discussed.
What is important here is that the reaction studied in the experiment of \cite{gom01} is
neither exactly within the scope of the original idea of Goldansky \cite{gol60}
for two-proton radioactivity, nor exactly within the scope of idea of
democratic decay \cite{boc84,dan87,boc92}. This implies that other
breakup mechanisms could also be important, and we now review such possibilities.

Qualitatively different cases of two-proton emission are illustrated in
Fig.\ \ref{sch}.
Presumably democratic two-proton decay of the $^{16}$Ne ground state
(Fig.\ \ref{sch}a, \cite{kek78,woo83,gri01b}), where there are no strongly
correlated subsystems,
is compared with decays of excited states in
$^{14}$O (Fig.\ \ref{sch}b, \cite{bai96}) and
$^{18}$Ne (Fig.\ \ref{sch}c, \cite{gom01}), where binary decay channels are
opened. The conceptual difference between the experiments Refs.\
\cite{bai96} and \cite{gom01}
is that there is a sequential decay branch through the 1/2$^+$ state at 2.37 MeV
in $^{13}$N (Fig.\ \ref{sch}b), which was shown to be important decay mode
in \cite{bai96}.
In the case of $^{18}$Ne sequential decay is not possible,
but one-proton decay is still allowed to the weakly bound states
in $^{17}$F (5/2$^+$ ground state, $E_b=0.6$ MeV and 1/2$^+$, $E_b=0.11$ MeV,
Fig.\ \ref{sch}c).
%They can easily break up under the Coulomb barrier providing important sequential
%mechanism for the decay.
In the recent experiment \cite{gom01} the flux
in the $^{17}$F+$p$ channel is orders of magnitude larger than in the three-body
$^{16}$O+$p$+$p$ channel and can not be neglected in the study of the $2p$
channel.
In such a case it will be very difficult to disentangle the three-body decay of
$^{18}$Ne states from the breakup of $^{17}$F on a proton target (which also
provides two protons in the final state), making the interpretation
of the reaction more complicated.

We therefore examine the nature of the  reaction  \cite{gom01} by
means of $R$-matrix theory, to see whether the observed two-proton events
can be explained in terms of three-body decay from the
$1^-$, $E^*=6.15$ MeV resonance state in $^{18}$Ne. We calculate the
two-proton production cross section and branching ratios.

{\em Estimates.} To estimate the cross section value for two-proton production
via the resonance compound state with definite $J^{\pi}$ we use the
standard formula (Ref.\ \cite{sat})
\begin{equation}
  \sigma^J_{\alpha \beta}(E) = \frac{\pi}{k^2} \,
  \frac{\Gamma _{\alpha}\Gamma _{\beta} }{(E_R - E)^2 + \Gamma ^2 / 4}\,
  \frac{2J+1}{(2J_{1 \alpha }+1)(2J_{2 \alpha}+1)} \,  ,
  \label{sig1}
\end{equation}
where $\alpha$ and $\beta$ label the entrance and exit channels,
$J_{1 \alpha}$ and $J_{2 \alpha}$ are the spins of the particles in the entrance
channel, $\Gamma$ is the total (experimental) width of the resonance
and $\Gamma_{i}$ are partial widths.
For the $1^-$ state $\Gamma=50$ keV \cite{gom01}.
To obtain the complete cross section in the case of elastic scattering
($\alpha = \beta$) the Coulomb and potential scattering together with
any interference terms must be added to Eq.\ (\ref{sig1}).

The resonant $2p$ production cross section via a $1^{-}$ state is
\begin{equation}
  \sigma^{1^-}_{2p}(E_R) = \frac{\pi}{k^2}\,
  \frac{\Gamma _{1p}(\mbox{gs}) }{\Gamma}\,\frac{\Gamma _{2p} }{\Gamma}\, ,
  \label{sig2}
\end{equation}
where,  evaluated on the resonance, $\pi/k^2=0.31$ barn, and $\Gamma_{1p}
(\mbox{gs})$ is the decay width to the $^{17}$F ground state.
It can be seen from (\ref{sig2}) that cross section value of 310 mb
is the upper limit for resonance processes via the $1^-$ state of $^{18}$Ne at
6.15 MeV.

If the decay of the $1^-$ state to the excited $1/2^+$ state
in $^{17}$F is negligible, the ratio $\Gamma _{1p}(\mbox{gs}) / \Gamma$
is very close to unity.
A more realistic  estimate takes into account
the variation of penetrabilities with decay energy.
The conventional R-matrix formula is \cite{lt}
\begin{equation}
\Gamma^{i}_{1p}(E) = 2\, S^{i}_{1p}\,\frac{3}{2 M^1_{17} r_c^2}\,
  P_l(E,r_c,Z(^{17}\mbox{F}))\, ,
\end{equation}
where $M^j_k$ is the reduced mass for particles with mass numbers $j$ and $k$. In
this article the channel radius $r_c$ is varied from 2.5 to 4.5 fm to give an idea
of the theoretical uncertainties associated with this parameter. Neglecting the
difference of spectroscopic factors $S^{i}_{1p}$ for the $^{17}$F(gs)+$p$ and
$^{17}$F($1/2^+$)+$p$ channels we obtain $\Gamma _{1p}(\mbox{gs})/ \Gamma
_{1p}(1/2^+) \sim 2$, and hence $\Gamma _{1p}(\mbox{gs})/ \Gamma  \sim 2/3$. If the
inelastic channel $^{17}$F($1/2^+$)+$p$ has a large spectroscopic factor, it can
influence drastically the cross section of the two-proton decay. This is connected
with the fact that the large spectroscopic factor $S_{1p}^{1/2+}$ will decrease the
ratio $\Gamma _{1p}(\mbox{gs})/ \Gamma $ in Eq.\ (\ref{sig2}) and will lead to
smaller $2p$ production through the $1^-$ resonance. Unfortunately there was no
experimental identification of the inelastic $^{17}$F($1/2^+$)+$p$ channel in
\cite{gom01}. Such identification is highly desirable to provide more reliable
estimates for the process.

%===============================================================================

\begin{figure*}[t]
\centerline{
        \psfig{figure=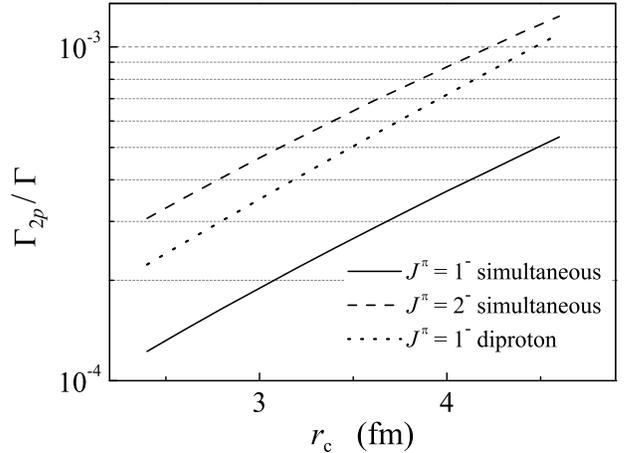,width=0.46\textwidth}
        }
\smallskip

\caption{The ratio $\Gamma_{2p}/\Gamma$ as function of channel radius $r_c$
estimates in two-proton and `diproton' models. Solid and dashed
curves correspond to $1^-$ state (simultaneous and `diproton' emission)
and dotted curve corresponds to $2^-$ state (simultaneous emission only).}

\label{cps}

\end{figure*}

To estimate the {\em simultaneous} two-proton emission width we
use a formula from \cite{gri01a} (similar to one given in \cite{gol60}):
\begin{eqnarray}
  \Gamma_{2p}(E_{2p}) = 2 \, S_{2p} \,
  \frac{3}{\pi r_c^3 (M^1_{16})^{3/2}E_{2p}^{1/2}}
  \int_0^{1}  dx
  \nonumber \\
  \times
  P_{l_1}\!\left(xE_{2p},r_c,Z(^{16}\mbox{O})\right)
  P_{l_2}\!\left( (1-x)E_{2p},r_c,Z(^{16}\mbox{O})\right) ,
  \label{sim1}
\end{eqnarray}
where $l_1$ and $l_2$ are the angular momenta of the 2 protons,
and  $E_{2p}$ is the three-body energy relative to the two-proton breakup threshold.
The coefficient in front of the penetrabilities
plays the role of a reduced width and is normalized in
the spirit of the Wigner limit: without any barriers the width is the average
inverse flight time for the distance $r_c/3$.
We certainly expect the two-proton spectroscopic factor to be smaller than the
one-proton spectroscopic factor: $S_{2p}<S_{1p}$.
However, assuming $S_{2p}=S_{1p}$ the upper limit for the estimated branching
ratio with $l_1=0$ and $l_2=1$ is $\Gamma_{2p}/ \Gamma =(1.5-5)\times 10^{-4}$
(for $r_c$ in the range 2.5--4.5 fm, see solid curve in Fig.\ \ref{cps}).

%===============================================================================
%===============================================================================

Another simple estimate of the branching ratio is provided by the
{\em diproton} model:
\begin{equation}
\Gamma_{2p}(E_{2p}) = 2 \, S_{2p} \, \frac{3}{2 M^2_{16} r_c^2 } \,
  P_l(E_{2p}-\epsilon,r_c,2Z(^{16}\mbox{O}))\, ,
\label{dipro}
\end{equation}
where $l=1$ for the $1^-$ state and the average internal energy of the
`diproton' $\epsilon \sim 0.3$ MeV is
estimated from the experimental distribution Fig.\ 4 in \cite{gom01}.
The usual assumption is to take $\epsilon = 0 $
(see for example \cite{bro91,naz96})
or $\epsilon = 0.05-0.1$ MeV (for example \cite{gol65,jan65})
as the energy of the `virtual state' in two-proton system, but these
prescriptions run into methodological problems \cite{bar01,gri01a}.
There is no need to follow these prescriptions if one has an idea what
this energy actually is.
The use of the `experimental' diproton energy
in Eq.\ (\ref{dipro}) provides an estimated branching ratio
$\Gamma_{2p}/ \Gamma =(2-10)\times 10^{-4}$
which is in a very good agreement with the simultaneous emission estimate
above.
Figure \ref{cps} shows the ratio $\Gamma_{2p}/\Gamma$ for both types of
estimates as a function of channel radius $r_c$.

%===============================================================================

{\em Discussion.}
In the estimates above we have made several assumptions.
Each of them is likely to give us the {\em upper} limit
for the $2p$ production cross section.
We should also note that the source of the theoretical uncertainty in the
2$p$ cross section is a variation of the channel radius $r_c$.
The source of experimental uncertainty in the two-proton production cross
section is connected with different assumptions about the decay process
(`diproton' decay or `democratic' decay) in the analysis of the data,
because of the limited acceptance of the experiment.

The estimates for $\Gamma_{2p}/ \Gamma $ ratio are
consistent with theoretical estimates in \cite{gom01}.
We deduce from Eq.\ (\ref{sig2}) an estimated  two-proton
production cross section through the  $1^-$ resonance of $0.03-0.2$ mb,
much smaller\footnote{
The branching ratio only gives the experimental cross section
if it multiplies a cross section of 3660 mb  \cite[in note 16]{gom01}
for 1$p$ resonance production from Eq.\ (\ref{sig2}).
}
than the measured value  $1.5-4.0$ mb reported in \cite{gom01}.

%===============================================================================
\begin{figure}[tb]
\centerline{
        \psfig{figure=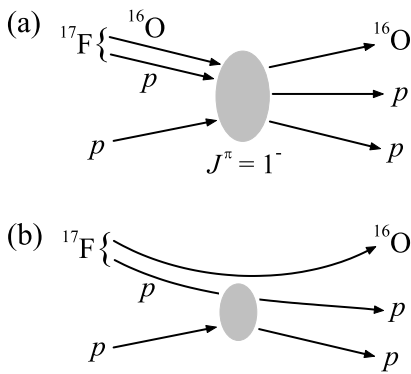,width=0.30\textwidth}
        }
\smallskip

\caption{Schematic presentation of (a) emission of two protons via the
decay of resonance in $^{18}$Ne, and (b) dominating mechanism of $^{17}$F
breakup on protons.}

\label{bru}

\end{figure}
%===============================================================================

There are two possible explanations of the large two-proton production
cross section seen in the experiment of \cite{gom01}.

(i) According to the level scheme in the isobaric $^{18}$O nucleus,
and to the experimental evidence from \cite{hah96,gom01},
the $2^-$ and $3^-$ states are located
only slightly higher than the $1^-$ state.
Two-proton contributions from the $3^-$ state should be negligible, as
$l_1=1$ and $l_2=2$
for the simultaneous emission model Eq.\ (\ref{sim1}), or $l=3$ for
`diproton' emission Eq.\ (\ref{dipro}).
The `diproton' emission from a $2^-$ state is parity forbidden as has
been mentioned in \cite{gom01}, but simultaneous two-proton emission
is allowed with the same quantum numbers $l_1=0$ and $l_2=1$ as from the
$1^-$ state, with a width comparable with the `diproton' mechanism
(see Fig.\ \ref{cps}, dashed line).
Using Eqs.\ (\ref{sig2}) and (\ref{sim1}) we obtain the value of
$\sigma_{2p}^{2^-} \sim 0.1-0.4$ mb.
This value would decrease
significantly the difference between measured and
estimated two-proton production cross sections.

%===============================================================================
\begin{figure}[tb]
\centerline{
        \psfig{figure=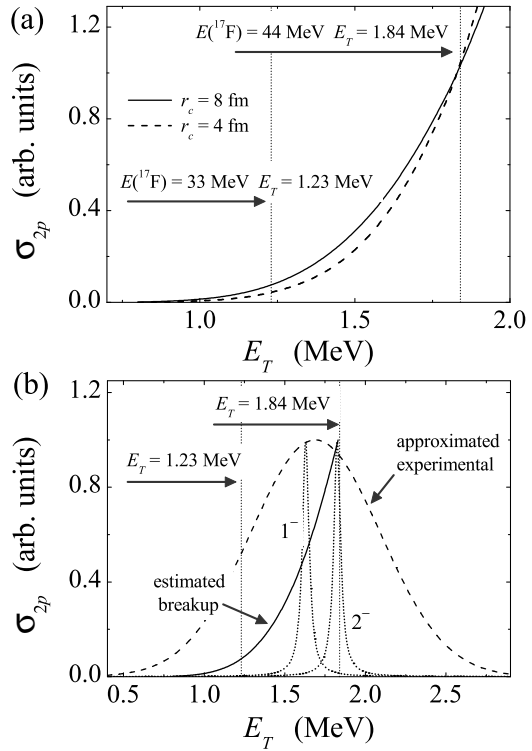,width=0.44\textwidth}
        }
\smallskip

\caption{Qualitative comparison of different possible contributions to the
two-proton events. (a) Shows the estimated contributions of possible breakup
events in experiments at 33 and 44 MeV beam energies (for different channel radii).
Fig.\ (b) compares the expected shape of the strength functions for breakup events
(solid curve is the same as in upper panel), resonance events
(from 1$^-$ and 2$^-$ states; dotted curve),
and the experimental distribution approximated by a
gaussian (dashed curve).}

\label{nrc}

\end{figure}
%===============================================================================

(ii) The other possibility is that most of the two-proton events are
actually coming from the breakup of $^{17}$F on protons, rather than decay
of resonances in $^{18}$Ne.
Figure \ref{bru} schematically outlines the difference between the two
processes. It is difficult to estimate the breakup process consistently,
as it can involve the complicated interplay of two-body and three-body
dynamics.
What is important here is that this contribution is not limited by the
form of Eq.\ (\ref{sig2}) which applies only for $2p$ production via a
resonance.

The energy dependence of the two-proton breakup channel can be roughly estimated by
the formula for simultaneous two-proton emission, Eq.\ (\ref{sim1}), as both protons
have to penetrate through the Coulomb barrier. However, the channel radius no longer
has a well defined meaning, although it is certainly expected to be larger than the
values used in estimates for the decay of a compound state.
Imagine for the moment that there is no resonance contribution
to the two-proton production at all. The estimated energy dependence of the
breakup cross section is shown in Fig.\ \ref{nrc}a.
In the type of experiment described in \cite{gom01}, all beam energies below maximal
are present with comparable probabilities because of the thick target used.
The estimated ratio of breakup events observed in experiments
with 44 MeV and with 33 MeV beams is 20--40 depending on channel radius.
This is given by the ratio of integrals of the intensity shown in Fig.\ \ref{nrc}a
up to $E_R=1.84$ MeV and $E_R=1.23$ MeV respectively.
The main contribution of two-proton events from inelastic
breakup comes from the maximal energies available in the experiment.
In Fig.\ \ref{nrc}b this contribution is qualitatively compared
with the expected contribution from $1^-$, $2^-$ states and the distribution
of actually observed events, which is broad due to the energy resolution
being low in the experiment. In our view all the above mechanisms could contribute to the
experimental cross sections.

The ratio of the
two-proton cross section for the $E_{lab}(^{17}$F$)=33$ MeV measurement
compared to the $E_{lab}(^{17}$F$)=44$ MeV measurement could be possible
evidence for identification of decays via the $1^-$ resonance,
and the experiment \cite{gom01} sees a 7--10 fold suppression at the lower energy.
However, we have to be careful that the growth of direct breakup does not
dominate any resonance contributions.
No resonant 2$p$ contribution is expected in the 33 MeV measurement.
The estimates of Fig.\ \ref{nrc} shows that if the number of two-proton
events in 33 MeV measurement is only 10 times lower than in 44 MeV
measurement, then extrapolation of the intensity of breakup to higher
beam energy gives more than enough 2$p$ events to  explain the whole $2p$
intensity in 44 MeV measurement. This indicates that further experimental
evidence will be need to discriminate between direct breakup and resonant decay.

{\em Conclusion.} We have shown that it is likely that the two-proton decay of the
$1^-$ state is sufficient to explain only a small fraction of the $2p$ events
reported in Ref.\ \cite{gom01}. It is plausible  that the balance of $2p$ events is
connected either with (i) excitation of  a $2^-$ state located a little higher than
the $1^-$ or/and with (ii) the nonresonant breakup of $^{17}$F on protons.
Identification of the inelastic breakup channel $^{17}$F$(1/2^+)+p$ for the states
involved in the two-proton emission is also desirable for a refined interpretation
of the data. A complete kinematics experiment would be the most useful in the study
of $2p$ emission from $^{18}$Ne. It would allow the energy of the decaying states in
$^{18}$Ne to be fixed, and would thus make interpretation of the the experiment
clearer.

{\em Acknowledgements.}
The authors are grateful to L.\ Chulkov, V.\ Goldberg, B.\ Jonson, A.\ Korsheninnikov,
I.\ Mukha, W.\ Nazarewicz, and G.\ Nyman for useful comments.
L.V.G.\ is grateful for support from the Royal Swedish Academy of
Science and hospitality of the Chalmers University of Technology,
where part of this work was done.
We acknowledge the support of EPSRC Grant GR/M/82141 and RFBR Grant 00-15-96590.

\end{document}